# Exploitation of the genomic double-strand breaks to reduce the reproductive power of microorganisms

G. Sassi [1,*], N. Sassi [2,+]

[1]Department of Physics, University of Milano Bicocca, Milano, Italy

[2]Department of Biomedical Sciences, University of Padova, Padova, Italy

[*]Corresponding author. E-mail: giandomenico.sassi@unimib.it

[+] Currently in a different position

## Abstract

It is shown how to take advantage of the frequent occurrence of double-strand breaks in the genome of prokaryotic cells, in order to reduce their high efficient reproductive capability. The analysis examines the physical status of the free ends of each break and considers how this status can interfere with an external physical apparatus, with the aim of undermining the repair processes. We indicate the biological consequences of this interaction and we give an approximate evaluation of the topological and dynamical effects that arise on the genomic material involved. The overall result suggests a significant reduction of the dynamics of the repair.

# Introduction

To address the problem of the infection control and fight against resistant microorganisms, we can proceed along a new road, which involves a purely physical status that frequently occurs in the phenomenology of the bacterial genome when exogenous or endogenous breaks of the helix occur. What will be shown is that it is possible to generate an interaction between this genomic status and a simple external physical apparatus, with the result of decreasing the remedial options of the damage. As will be easily understood in what follows, only extensive testing can substantiate, or not, the goodness of the proposal. With these notes we just want to give credibility elements that justify the large experimental effort that must be undertaken. Moreover, even the alarming expectations connected to the use of antibiotics are involved, because below we will show that the present proposal, besides introducing a new perspective, can also increase antibiotics efficacy.

With reference to the phenomenology, we know that a break alters the spatial configuration of the double helix, and this topological change pushes the cell to bind a complex of specific molecules to the site of damage, so that the repair can take place. The physical effect that we are proposing acts on the dynamics of the damage, amplifies the phenomenon and turns away the ends of the DNA breaks, thus making more complicated the process of repair. A possible outcome of this amplification of the morphology of the damage is either the death of the cell, unable to repair the damage itself, or a significant delay of the repair processes.

All this can be achieved by an interaction between a magnetic field gradient and some specific atomic magnetic moments of the genome, the presence of which, to our knowledge, has not been conveniently exploited. Unlike the magnetic field gradients used, for example, in the nuclear magnetic resonance in order to interact with nuclear magnetic moments, we now consider a static field gradient with the aim of exerting forces on atomic magnetic moments which are generated in particular conditions. These conditions occur in critical circumstances for the bacteria and then these forces are particularly efficient.

We intend to make clear, and we will return on this, that the phenomenology we are presenting involves not only bacteria, but also the cells of the body. We will show the reasons why it can be assumed that the cellular damage is much lower than the effects on bacteria. Obviously, it will be task of the experimentation to confirm, or not, these predictions. At the moment we simply can observe that, if we pose the question of how relevant for the cells is the slowing of the repair process, cells that divide more rapidly, in particular prokaryotic cells, are preferably affected. In these bacterial systems, the evolutionary demand for rapid growth is particularly acute, and then the repair processes are so well coordinated that cells normally have success in overcoming the related problems. E. coli, for example, reproduces its entire genome in less than the 40 minutes required for its duplication(*1*). The physical mechanism that we are considering alters this solving power of the damage.

## Physical aspects

When the DNA strands break either for environmental reasons or for biological processes, the particular aspect that we must consider is that on the two ends (Fig. 1A) may be present non-zero magnetic moments. We can briefly remember that the magnetic moment of an atom is a function of the spins and orbital angular momenta of its electrons, whereas the contribution of the nuclear angular momentum is negligible. However, unless in some extreme conditions, it is not necessary to consider all the contributions of the individual electrons, because we can assume, without committing significant errors, that the magnitude of the resultant magnetic moment of an atom is either zero or approximately equal to the *Bohr magneton,*

$$\mu_B = \frac{q\hbar}{2m} = 9.27 \times 10^{-24} A \cdot m^2$$

which is very close to the magnetic moment of the spin of a single electron.

Not all atoms have permanent magnetic moment. If the angular momenta compensate, there is no resultant magnetic moment. A permanent magnetic moment is present, for example, in free radicals, because of their odd number of valence electrons, whose bonds are not fully saturated. When molecules are considered, usually electrons in the outside shell of an atom couple with electrons in the outside shell of another atom, with their spins in exactly opposite directions, as required by Pauli's exclusion principle. Consequently, the angular momenta and magnetic moments of the valence electrons are normally cleared. This is the reason why the molecules, as a rule, do not have magnetic moment.

However, when a molecule without net magnetic moment breaks -as in the case of the double strand of DNA- at both ends of the break may be present permanent uncompensated magnetic moments of equal value and opposite sign. This may occur with relevant probability, because two atoms that previously compensated their magnetic moments can

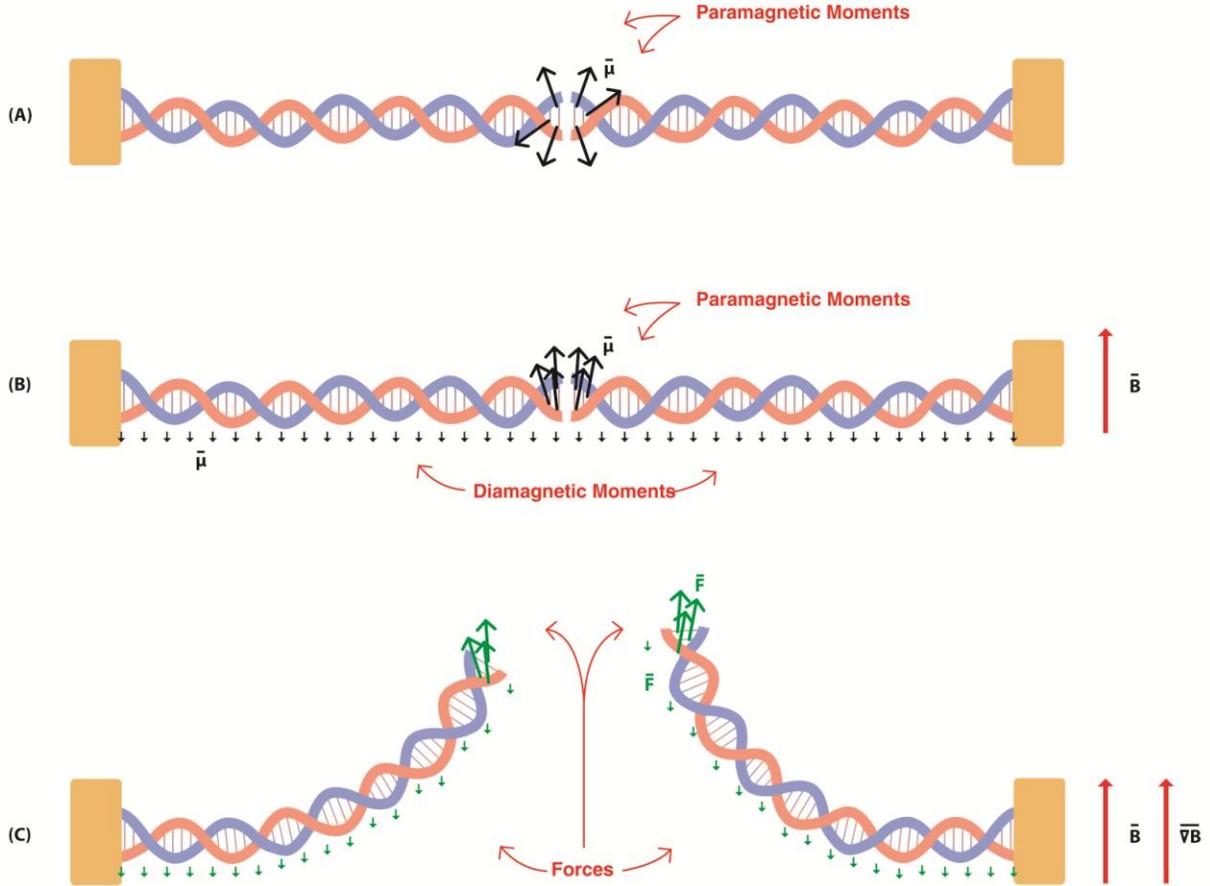

**Fig. 1. Double strand break behavior in presence of a magnetic field gradient.**

(**A**) Both exogenous and endogenous DSBs can possess paramagnetic free ends due to the presence of magnetic moments not saturated, originated by the rupture of the covalent bonds. On the two free ends the different magnetic moments must have equal values and opposite directions. (**B**) In the presence of an external magnetic field $\vec{B}$ there is a partial alignment of the magnetic moments along the direction of $\vec{B}$. Small magnetic moments of diamagnetic nature, and directed in opposition to $\vec{B}$, also arise along all the base pairs of the two cantilevers. The free ends are neither topologically nor dynamically affected by all these types of magnetic moments. (**C**) If the external magnetic field $\vec{B}$ has a gradient, $\overrightarrow{\nabla B}$, the magnetic moments give rise to forces. The topological and dynamical nature of the free ends changes completely.

now be placed separately on the two free ends. Fig.1A shows some magnetic moments $\vec{\mu_B}$ of this type, assuming that more atoms are involved. When this happens, we can say that the breakup leaves two paramagnetic free ends.

If an external magnetic field $\vec{B}$ is acting on the region involved in an infection, for all the double strand breaks (DSBs) that occasionally occur in the related bacteria, the result is that the magnetic moments $\vec{\mu_B}$ present on the ends of the breaks will be partially aligned in the direction of $\vec{B}$ (Fig. 1B). In fact, the presence of $\vec{B}$ gives rise to the energy $-\vec{\mu_B} \cdot \vec{B}$, which is a function of the angle θ between $\vec{\mu_B}$ and $\vec{B}$, and the statistical mechanics (as well as quantum mechanics) tells us that the probability that $\vec{\mu_B}$ forms a given angle with respect to $\vec{B}$ is given by $e^{-energy/KT}$. So with $\theta = 0$, the energy is minimum and the probability is maximum. The result is that on the free ends of the DSB a magnetization $\vec{M}$ arises, directed as $\vec{B}$ and with module (2)

$$M \cong \frac{N\mu_B^2 B}{3KT}$$

where $N$ is the number of atoms, belonging to each free end, that after the break possess the magnetic moment $\vec{\mu_B}$.

It is possible to think that a situation of this kind is able to complicate the possible re-composition of the break. In fact, the subsequent recombination of the two ends of the break is only possible if the electrons, which have to reform the covalent bonds, have opposite spins. The $\vec{\mu_B}$ realigned as in Fig.1B are in fact related to the spins of these electrons and being not oppositely directed the reconstruction is hampered. Actually, in ref. (*3*) and (*4*) no genetic effects has been detected with a strong static magnetic field (6.34T) even for a long exposition of 24h. In particular were not recorded significant microsatellite changes in both mismatch repair deficient and proficient cells. In ref. (*5*) it is instead suggested that, in the presence of static fields, each welding of genomic material is made more difficult by the consequent realignment of the magnetic moments, based on the exclusion principle. This uncertainty is not very relevant for the present analysis, because it is only with what is shown in Fig.1C that the phenomenon, as we shall see, is important. Of course, the chaotic thermal motions to which these ends are subjected imply that the scheme in Fig.1B represents only the average situation. So, the re-composition, if any, is obviously always possible, but it is a bit more difficult.

In Fig.1B are also present small magnetic moments opposed to the external field $\vec{B}$ and distributed on all DNA bases, so even on the bases that compose the cantilevers originated by the rupture. They represent the diamagnetic contributions that matter always shows in the presence of an external magnetic field. Generally, these magnetic moments can be neglected due to their small intensity. However, in this case we have to consider that the aromatic groups have a significant diamagnetism perpendicular to their plane, and since the DNA contains conjugated bonds in its nucleic acid bases, it has been suggested that even DNA can have a similar behavior in the presence of $\vec{B}$ (6). In particular, it was established that the magnetic susceptibility, $\chi_{mol}$, of DNA, mediated on different bases and different orientations, gives a value of the order of $-10^{-4}\ emu/mol$. Since the quantity $\frac{B}{\mu_0}\chi_{mol}$, where $\mu_0$ is the magnetic permeability in vacuum, has the meaning of total magnetic moment present on one mole of DNA under the influence of an external field $\vec{B}$, we can easily calculate that even with a considerable field $\vec{B}$, say 10 T, the magnetic moment on the single base is $\sim 1.5 \times 10^{-23} A \cdot cm^2/bp \cong 10^{-4}\mu_B$. So we can confirm that the diamagnetic contributions are still much lower than the paramagnetic values present on the ends of the break.

Of course, both paramagnetic and diamagnetic moments are not forces and thus their presence does not alter the conformation of the DSB, either from the topological point of view or from the kinetic one. Changes of this type exist, however, if the external field $\vec{B}$ has a gradient $\vec{\nabla B}$, for example along the same direction in which $\vec{B}$ is oriented. In that case the magnetic moments originate forces that act on the free ends of the DSB, causing a lifting and therefore a mutual spacing, as shown in Fig.1C. In fact, in the presence of a magnetic field gradient, the magnetic energy $U = -\vec{\mu_B} \cdot \vec{B}$ varies with the position, and therefore a force on the magnetic moment arises, with module (2)

$$F_\mu = \mu_B \cos\theta \frac{\partial B}{\partial z}$$

Where $|\vec{\nabla B}| = \frac{\partial B}{\partial z}$ and z is the direction of $\vec{\nabla B}$. Values of $\vec{\nabla B}$ easily obtainable and sufficiently intense, range from $1 T/m$ to $10 T/m$, and therefore we will assume these values for our next evaluations.

### Biological effects

In the next section we will describe how the free ends can move under the effect of these forces. We will assess the extent of these movements and the time required. But now let us consider the expected biological effectiveness of the widening shown in Fig.1C, i.e. how this enlargement can reduce the reproductive performance of a microorganism.

First of all, it has been shown that the nucleoids of some bacteria (E. coli and B. subtilis) become significantly more compact after damaging conditions (*7-10*). Due to the nucleoid excision repair protein $U_{vr}A$, this reconfiguration of the nucleoid appears as a controlled response to DNA damage and may be the indication of ongoing repair. Apparently, keep as close as possible the damaged parts helps protein complexes that catalyze the repair processes. It follows that we are entitled to assume that the configuration shown in Fig.1C represents a plausible stratagem to reduce or slow down the ability of the bacterial cell of mending the broken double helix.

With reference to the phenomenology of these damages, we have to remember that the repair of DSBs can be obtained by the process of homologous recombination (HR), in which the breaking of the duplex is recomposed through a sequence of copies of a homologous chromosome, usually the sister chromatid, which functions as a template . The repair can also occur through non-homologous end-joining (NHEJ), in which many small insertions are made at the site of the damage, so as to obtain the bridging and ligation of the two free ends. During these processes, the ends of the broken DNA molecule must move and invade the template, or move to find themselves close to each other. There is a significant probability that even these various free ends can possess magnetic moments, and the proposed mechanism acts on the latter and interferes with the natural movements of these insertions. In the NHEJ, rejoining occurs through many small homologous sequences (1-6 bp), while in HR the repair is obtained with sequences of homology much longer, of the order of 100 bp(*11*). The presence of these new forces will tend to deform and reorient these sequences during their migrations, thus behaving as an obstacle.

We must also remember that DSBs are typical conditions for the start of the process of genetic recombination. Thus, a fundamental process such as recombination requires a potentially lethal condition. The phenomenon we are considering acts on it, although this is not a classical damage, causing thus actually a damage.

Effects are also expected from the action of topoisomerases, that cause single strand breaks in order to change the state of supercoiling of the DNA. Again, though these are not damages and there are enzymes responsible for a rapid repair, the forces of Fig.1C, although on a single strand, will cause harmful effects due to the divergence of the ends.

In general, the genome, particularly that of bacteria, is continually subjected to cuts and rejoinings. In all these events, the divergence shown in Fig.1C has as allies the exonucleases present in the cell . Since the re-composition of each break is made more difficult, and then slowed down, it is left more time to exonucleases to attack the free ends, and cause lethal damage to the genome.

One may also wonder whether this slowdown of the break repair interferes with the necessary coordination of the synthesis processes involved in cell division. In fact, the model with which to describe the bacterial division cycle includes three types of

patterns(*12*): the synthesis of the cytoplasm, of the genome, and of the cell surface. There are rules that link these three categories during the division cycle. In essence, the timings of these three different syntheses are tuned in order to maximize the effectiveness of cell duplication and to avoid the death of the cell. Being the concatenation so critical, a temporal alteration of the biosynthesis of the genome counteracts the required synchronicity of these processes. In short, the cell may not have all the needed genomic material at the time of its division, with inevitable negative effects.

Finally, coming back to exogenous DSBs, we must remember that among these there are breaks caused on bacterial genome by the traditional therapeutic treatments, as can happen with antibiotics. The process we are considering can enhance their effectiveness, because the two different treatments can feed on each other, as one creates damage that the other amplifies, and they could be optimized in order to create a sort of therapeutic resonance. To mention a particular example of this kind, we can consider the case of the bactericidal function of quinolones. Their effect of trapping gyrase and topoisomerase IV on DNA leads to the release of DSBs (*13,14*). DNA fragmentation can arise and many DNA ends can lead to cell death. However, this clinical result often is not achieved because of an accumulation of resistance mutations. The divergence of the free ends here considered, amplifying the damage, can prevent mutations, because the latter, in order to happen, necessarily require that the welding occurs.

In conclusion, it is worth remembering that the complexity of the biochemical processes that take place in each cell is so high that many other effects, perhaps more significant than those here described, may occur in relation to the presence of the phenomenology shown in Fig.1C. The scenario is necessarily broad and engaging.

### Dynamics of the process

We have to consider how the free ends of a DSB can migrate and move away, so as to perform the divergence shown. Their movement depends primarily on the extension between each free end and the first anchor point, i.e. the extension of the portion of double helix that assumes the nature of a cantilever. In fact, it has been proved the existence of meeting points between different locations of bacterial chromosomes (*15,16*). These contact points, shown in Fig.1 with generic blocks, give rise to a natural periodicity and consequently they create different neighborhoods along the genome. Each DSB necessarily occurs between two of these meeting points, thus defining the two cantilevers.

In conjunction with the extension of the cantilevers it is necessary to consider the bend-persistence length of the double helix, which is the length scale beyond which the cost of the elastic bending is totally negligible(*17*). For DNA this length is estimated to be $50\ nm$, i.e. about $150\ bp$. For simplicity, we evaluate the raising of the two free ends in Fig.1C assuming that the cantilevers have just extensions equivalent to the bend-persistence length. If their lengths were smaller (larger), the increase would be lower (larger), respectively. For this purpose we introduce the wormlike chain model that appropriately describes DNA as a chain of subunits joined in a partially flexible hinges, because the different segments, consisting of base pairs, display a sort of cooperativity (*18*). In essence, we have to consider how the free ends may rise up after the rupture without implying an energy cost, and therefore without supposing that there is an elastic increase in the length of the cantilevers, so that no entropic elasticity is involved. In fact, the force acting on each free end is too small to be able to develop sufficient energy to reduce the degrees of freedom of the chain.

In the zero stretching force approximation, i.e. with a very small force, the tangent-tangent correlation function for the wormlike chain is given by(*18*)

$$\langle \hat{t}(0) \cdot \hat{t}(s) \rangle = e^{-|s|/\xi}$$

where $\hat{t}(0)$ is the unit tangent vector to the chain in the meeting point where the cantilever is anchored, $\hat{t}(s)$ is the unit tangent vector to the chain on the free end where acts the paramagnetic force, $s$ is the length of the chain between these two points and $\xi$ is the bend-persistence length. The meaning of this relationship is that the curvature that this portion of the chain can assume at zero (or very low) stretching force is such that the cosine of the angle formed by the two unit tangent vectors is, on average, given by $1/e$ if the length $s$ of the cantilever is exactly equal to the bend-persistence length $\xi$. In our case, assuming that the cantilever has a length of $50\ nm$, the angle between the two unit vectors turns out to be, on average, $68°$. Given the angle, in order to know precisely the extent of the raising or lowering of the free ends, and consequently to know the amount of their moving away from each other, we should know the shape of the curve that the two cantilevers take on. Since this is unknown, we can safely assume that with $s = \xi = 50\ nm$ and the relative angle of $68°$, the lifting of the ends can reasonably be $20 \div 30\ nm$ and the their separation can be about $5 \div 10\ nm$. The extent of this separation is certainly significant. Of course these movements are added to the chaotic thermal motion. In other words the two cantilevers will continue to oscillate, towards and away from each other due to thermal motions, but their average position will be in this condition of curvature and not in the alignment possessed at the time of rupture.

We need at this point to estimate the time required by the free ends to acquire this divergence after the occurrence of the DSB. If the divergence must serve to make it more difficult the re-composition of the break, the time required must be competitive with those associated with the very rapid genomic adjustments. Inside a prokaryotic cell, DNA represents 2% of the total mass (*19*). So, inside the nucleoid of the cell we can assume that DNA represents about 10% of the mass of the nucleoid itself. If we consider the molecules of water associated with the double helix and the various nucleoid associated proteins (NAPs), all these compacted masses can get close to half the mass of the nucleoid and have to move together. The consequence is that the movements of the free ends of a DSB cannot be described as viscous motions of a particle in a fluid, being the mobile mass comparable with the neighboring mass. So we must renounce to the convenient description of these movements on the basis of the concepts of viscosity, diffusion tensor and vector drift. Similarly, it is not through the Fokker-Planck equation (*20*) that we can establish the distribution of velocities of these free ends, and consequently the extension of their displacement and the time that this requires. Instead, it is more reasonable to assume, for simplicity, that the cantilevers are moving dragging with them a mass 5 times greater, just to say, than that of the genome involved, with a drift motion limited by generic impacts with other material present in the nucleoid.

Now we can share a numerical evaluation, necessarily approximate and only indicative. As mentioned, we assume that the cantilevers extension is equivalent to a bend-persistence length, that is $150\ bp$. The double helix corresponding to this length has a mass of about $90\ KDa$ and, considering the aggregated masses, we assume that the total mass that has to move can be established in $450\ KDa \cong 0.75 \times 10^{-21} Kg$. At room temperature each cantilever oscillates with a thermal velocity $v_{th} = \sqrt{3KT/M} \cong 1.5\ m/s$. For these oscillations we expect that the condition, $\langle x \rangle = 0$, is satisfied, the same one that we should assume for a diffusion process. In other words, the cantilever vibrates around a point of equilibrium which depends on the position of the DSB at the time of breakage, but also on the tensions present in the genome itself and that can change its position. If we assume that at the end of each cantilever there is the presence of a single magnetic moment $\vec{\mu_B}$, immersed in an external magnetic field gradient $|\vec{\nabla B}| = 10$ T/m, easily obtainable, the force on each free end would be $F \cong 10^{-22}\ N$, in the direction of $\vec{\nabla B}$. At this point we can make many assumptions on the effect of this force. Imagine, for example, that the end of the cantilever can move freely and acquire velocity, due to this force, only for a distance of $0.1\ nm$, which is $1/20$ of the diameter of the double helix, and from that moment on it maintains the velocity assumed, without increasing it, as a consequence of the various impacts. The velocity would be $v_d = \sqrt{2lF/M}$, and then with $l = 10^{-10} m$ we have $v_d \cong 5 \times 10^{-6} m/s$. With this drift velocity maintained throughout the displacement of $30\ nm$, the time required for the total movement is about $6 \times 10^{-3} s$. If during this lapse of time the soldering of the DSB has not yet been obtained, its subsequent achievement will result more difficult.

### Final notes

It is licit to object that the calculations made in the previous section are very approximated and therefore the movements of the free ends may be very different. Roughly, we can nonetheless still say that the divergence of the free ends takes place in a time comparable to that of the biological processes in which it is involved. Of course some DSBs can happen in such a way (or under such conditions) to reduce dramatically the divergence of the free ends, or even prevent it completely. But it should be remembered that there are many occasions in which the DSB's occur, and even though only a few of these manifest the displacement expected, the probability of bacterial duplication can be significantly reduced. Similarly, even if the effect here considered would concern only a portion of the bacteria involved, a significant therapeutic effect may still occur.

Coming back to the consequences of this phenomenon on the cells of the body, two reasons lead us to believe that they are vastly inferior to those obtainable on the bacterial cells. First of all because, unlike the latter, their duplication is less frequent. Then because, even in the presence of duplication, this requires much longer times, typically 24 hours. If we assume that the field gradient we are considering is made active only for a time corresponding to that of duplication of the bacteria, perhaps repeated a few times during the 24 hours, the total time of the activity would be a small fraction of the duplication time of human cells. Of course, among the latter there are also cells which are more rapid in their reproduction and, in general, generated precisely to counteract an infection. We have, however, the advantage that their generation takes place in parts of the body that, very often, are different from those in which the field gradient is applied, which is the region involved in the infection. In case of sepsis, or generally with the presence of bacteria in the blood, an extracorporeal circulation should be chosen as a site subjected to the action of this procedure.

Finally, a non-technical comment. Since the presence of a magnetic field gradient in concurrence with the formation of a DSB is not a common situation in nature, the bacterial evolution did not elaborate solutions that can cope with the problem, and this bacterial difficulty necessarily will continue for a long time. The need for something unusual, not evolutionarily recognized by bacterial *wisdom*, is the crucial requirement that every new attempt must possess.